\documentstyle[sprocl,epsf]{article}

\def\be{\begin{equation}}
\def\ee{\end{equation}}
\def\bea{\begin{eqnarray}}
\def\eea{\end{eqnarray}}
\def\hal{{1\over 2}}
\def\bx{{\bf x}}

\begin{document}

\title{SCREENING MASSES OF HOT SU(2) GAUGE THEORY FROM THE 3D
ADJOINT HIGGS MODEL}

\author{F. Karsch}

\address{Fakult\"at f\"ur Physik, Universit\"at Bielefeld,\\
\qquad  P.O. Box 100131, D-33501 Bielefeld, Germany\\
E-mail: karsch@Physik.Uni-Bielefeld.DE}

\author{M. Oevers}

\address{
Department of Physics and Astronomy, University of
Glasgow,\\
Glasgow,  G12 8QQ, U.K. \\
E-mail: m.oevers@physics.gla.ac.uk}

\author{ P. Petreczky}

\address{
Department for Atomic Physics, E\"otv\"os University,\\
Pazm\'any P\'eter s\'et\'any 1, 1117 Budapest, Hungary\\
E-mail: petr@cleopatra.elte.hu}


\maketitle\abstracts{
We study the Landau gauge propagators of the lattice 
$SU(2)$ 3d adjoint Higgs
model, considered as an effective theory of high temperature 4d $SU(2)$
gauge
theory. From the long distance behaviour of the propagators we extract
the screening masses. It is shown that the pole
masses extracted from the propagators  agree 
well with the screening masses obtained recently in finite
temperature $SU(2)$ theory. The relation of the
propagator masses to the masses extracted from gauge invariant
correlators is also
discussed.
In so-called $\lambda$ gauges non-perturbative evidence is given for
the gauge independence of pole masses within this class of gauges.
}


\section{Introduction}
The screening of static chromo-electric fields is  one of the most outstanding
properties of $QCD$ and its investigation is important both from
a theoretical and phenomenological point of view (for phenomenological
applications see e.g.~\cite{wang}).  In leading order of perturbation
theory the associated inverse screening length (Debye mass) is defined
as the $IR$ limit of the longitudinal part of the gluon self energy
$\Pi (k_0=0,{\bf k} \rightarrow 0)$. However, as the screening
phenomenon is related to the long distance behaviour of $QCD$ 
the naive perturbative definition of the Debye mass
is obstructed by severe $IR$ divergences of
thermal field theory and beyond leading order the above definition
is no longer applicable.
Rebhan has shown that the definition of 
the Debye mass through the pole of the longitudinal part of the gluon
propagator is gauge invariant \cite{rebhan}. 
However, this definition requires the introduction of a so-called
magnetic screening mass, a concept introduced long ago \cite{linde}
to cure the $IR$ singularities of finite temperature non-Abelian theories.
Analogously to the electric (Debye) mass the magnetic mass can be
defined as a pole of the transverse part of the finite temperature 
gluon propagator. 

As the screening masses are static quantities it is expected that they
can be determined in
a 3d effective theory of $QCD$, the 3d $SU(3)$ adjoint Higgs model, provided
the temperature is high enough. However, in the case of $QCD$ one may
worry whether the standard arguments of  dimensional reduction
apply because the coupling constant is large $g \sim 1$  
for any physically interesting temperature and thus the requirement
$g T << \pi T$ is not really satisfied. 

The main question which we will try to clarify in this contribution is
whether the screening masses,
defined as  poles of the corresponding lattice propagators in 
Landau gauge, can be determined in the effective theory for the simplest
case of the $SU(2)$ gauge group, where precise 4d data on  screening
masses  exist for a huge temperature range \cite{heller}.
We will also study the screening masses in the so-called
$\lambda$-gauges in order to test non-perturbatively the gauge
dependence of the pole masses which were proven to be gauge invariant
in perturbation theory \cite{kobes}.
Finally we will briefly discuss the connection between propagator pole
masses and the masses extracted from gauge invariant correlators.

\section{Numerical Results on the Propagator Pole Masses}
The lattice action for the 3d adjoint Higgs model used in the present paper
has the form
\bea
&&
S=\beta \sum_P \hal Tr U_P + 
\beta \sum_{\bx,\hat i} \hal Tr A_0(\bx) U_i(\bx) A_0(\bx+\hat i)
U_i^{\dagger}(\bx) + \nonumber\\
&&
\sum_{\bx} \left[-\beta\left(3+\hal h\right) \hal Tr A_0^2(\bx) + 
\beta x { \left( \hal Tr
A_0^2(\bx)\right)}^2 \right],
\label{act}
\eea
where $U_P$ is the plaquette, $U_i$ are the usual link variables and
the adjoint Higgs field is parameterized 
by anti-hermitian matrices $A_0=i \sum_a \sigma^a
A_0^a$ ($\sigma^a$ {are the usual Pauli matrices}). Furthermore 
$\beta$ is the lattice gauge coupling, $x$ parameterizes the 
quartic self coupling of the Higgs field and $h$ denotes the
bare Higgs mass squared. This model is known to have two phases the
broken (Higgs) phase and the symmetric (confinement) phase separated by
the line of $1^{st}$ order transition \cite{nadkarni,hart,kajantie1}. 
The high temperature phase 
of the 4d $SU(2)$ gauge theory corresponds to some surface in the
parameter space $(\beta, h, x)$, the surface of 4d physics
$h=h_{4d}(x,\beta)$. This surface may lie in the symmetric phase or in
the broken phase, i.e. the physical phase might be either the symmetric
or the broken phase. In the context of a dimensional reduction performed
perturbatively at 2-loop level \cite{kajantie1} one finds 
the surface of 4d physics to lie in the broken phase. This, however, leads
to conceptual problems for the combined use of perturbation theory and
dimensional reduction because 
the expectation value of $A_0$ would then be ${\cal O} (1/g)$ for $g\ll 1$
although dimensional reduction is only valid if $A_0 \ll \pi T$. 
Still the surface of 4d physics determined 
perturbatively may be used in numerical calculations as this surface lies 
close to the $1^{st}$ order transition line. As the transition is in fact
strongly first order numerical simulations on finite lattice will be
performed in the metastable region and one can do simulations on the
symmetric branch in this region~\cite{kajantie1}.
The obvious problem with this approach is that the metastable region will 
disappear in the infinite volume limit. We therefore have performed  
simulations in the symmetric phase and determine the surface of 4d physics by 
non-perturbative matching \cite{ours} of a physical observable -- the gluon
screening masses.

We are going to review here our results on electric and
magnetic screening masses obtained from the Landau gauge propagators in
the symmetric phase as well as in the metastable 
region of our finite lattices.

Most of our numerical studies have been performed on lattices of size
$32^2\times 64$ and at $\beta=16$. 
The two sets of values of $h$ and $x$ used in our simulations in the
symmetric phase are shown in Table 1, where also the corresponding
temperature values as well as the values of $h$ corresponding to the 
transition are indicated. The temperature scale is
essentially fixed by $x$. The detailed procedure of choosing the
parameters in the symmetric phase is given in Ref.~\cite{ours}.
Simulations in the metastable region have been performed at the
values of the parameters obtained from 2-loop dimensional reduction 
\cite{kajantie1}. The results on the screening masses are summarized in
Figure 1 where also the results of 4d simulations \cite{heller} are
shown. As one can see from the figure the agreement between the masses
obtained from 4d and 3d simulation is rather good. The magnetic mass
practically shows no dependence on $h$ and its value is rather close to
the magnetic mass of 3d pure gauge theory $m_T=0.46(3)g_3^2$ ($g_3^2$ is
the 3d gauge coupling). The electric mass shows
some dependence on $h$, but with the present statistics all sets of $h$ 
values are compatible with the 4d data.
\vskip0.1truecm
\begin{center}
{\small
\begin{tabular}{|l|l|l|}
\hline
$~~~~{\rm Temperature~scale}~~~~$ &$~~~~~~~~~~~~~~~~~~~~~~~~~~~~~~~h~~~~~~~~~~~~~~~~~~~~~~~~~~~~$ \\
\end{tabular}
\begin{tabular}{|l|l|l|l|l|}
\hline 
$~~~~~x~~~~~$  &$~~~~~T/T_c~~~~~$   &$~~~~~~~I~~~~~~~$  &$~~~~~~~II~~~~~~~$ 
&$~~~~{\rm transition}~~~~$\\
\hline
$~~0.09~~$  &$~~~4.433$   &$~~-0.2652$   &$~~-0.2622$   &$~~-0.2672(4)$\\
$~~0.07~~$  &$~~~12.57$   &$~~-0.2528$   &$~~-0.2490$   &$~~-0.2553(5)$\\
$~~0.05~~$  &$~~~86.36$   &$~~-0.2365$   &$~~-0.2314$   &$~~-0.2399(6)$\\
$~~0.03~~$  &$~~~8761$   &$~~-0.2085$   &$~~-0.2006$   &$~~-0.2138(9)$\\
\hline
\end{tabular}
\vskip0.2truecm
Table 1: {\small The two sets of the
bare mass squared used in the simulation and those
which correspond to the transition line for $\beta=16$}
}
\end{center}
Let us turn to the discussion of the gauge dependence of the propagator
masses. To study the gauge dependence of the pole masses we have used
the so-called $\lambda$-gauges \cite{bernard} defined by the gauge 
fixing condition 
\be
\lambda \partial_3 A_3+\partial_2 A_2+\partial_1 A_1=0.
\ee
The case $\lambda=1$ corresponds to to the Landau gauge. 
We have measured the electric and the magnetic screening masses on
$32^2\times 96$ lattice at $\beta=16$, $x=0.03$ and $h=-0.2085$ for
$\lambda=0.5, ~1.0$ and $2.0$. The results of these measurements are
shown in Table 2. 
\vskip0.2truecm
{\small
\begin{center}
\begin{tabular}{|l|l|l|l|}
\hline
$~~~\lambda~~~$ &$~~~0.5~~~~$ &$~~~1.0~~~$ &$~~~2.0~~~$\\
\hline
$~~~m_D/T~~~$ &$1.14(12)$ &$1.11(7)$ &$1.06(12)$\\
\hline
$~~~m_T/T~~~$ &$0.36(9)$ &$0.39(7)$  &$0.36(4)$\\
\hline
\end{tabular}
\vskip0.2truecm
Table 2: The screening masses in $\lambda$ gauges.
\end{center}
}
\noindent
As one can see from the table with the present statistics no gauge
dependence can be observed.
\begin{figure}
\vspace{-1cm}
\epsfysize=6cm
\epsfxsize=8cm
\centerline{\epsffile{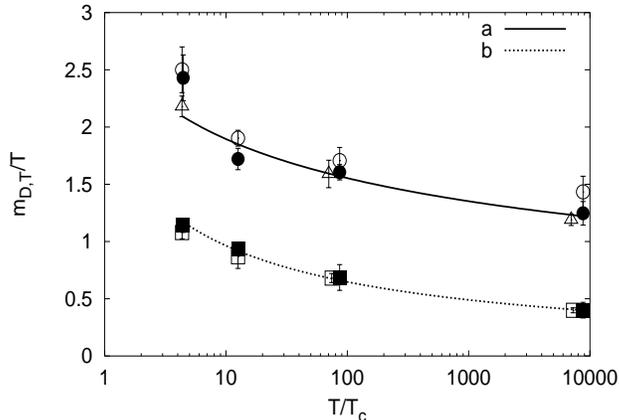}}
\vspace{-0.7cm}
\caption{The screening masses in units of the temperature. 
Shown are the Debye mass
$m_D$ for the first (filled circles) and the second (open circles)
set of $h$, and
the magnetic mass  $m_T$ for the first (filled squares)
and the second (open squares) set of $h$.
The lines (a) and (b) represent fits for the temperature
dependence of the Debye and the magnetic mass from 4d
simulations.
The open triangles are the values of the Debye mass calculated in the
metastable region by using
the coupling $h_{4d}(x,\beta)$ obtained from the 2-loop dimensional reduction.
Some data points at the temperature $T \sim 90 T_c$
and $T \sim 9000 T_c$ have been shifted in the temperature scale for
better visibility.}
\end{figure}
\section{Gauge Invariant Correlators and the Constituent Model}
Gauge invariant correlators for the SU(2) adjoint Higgs model were
studied in detail in Refs. \cite{kajantie1,philipsen1}. The masses
extracted from these correlators correspond to the masses of some
bound states. For example, the large distance behaviour of the
correlation function of the operator $Tr A_0^2$ yields the mass
$m(A_0)$ of the $A_0-A_0$ scalar bound state, 
the correlator of $h_{ij}=
Tr A_0 F_{ij}$ yields the mass 
of the bound state of the scalar field and light glue.
The masses of these bound states in terms of the constituent picture
are $m(A_0)=2 m_D$ and $m_h=m_D+m_T$ \cite{philipsen2}. 
The predictions of the
constituent model for the mass of the scalar bound state compared with
the results of the direct measurements are shown in Table 3. 
As one can see from the table the predictions of the constituent model
agree quite well with the measured data. 
For the mass of the bound state of the scalar field and the light glue the
constituent picture yields $m_h=1.83(4)g_3^2$ ( at $\beta=9$, 
$x=0.09$ and $h=-0.2883$ )
which should be compared with 
the result of the direct measurement $m_h=2.16(20)g_3^2$
from Ref.
\cite{philipsen1}.
\vskip0.2truecm
\begin{center}
{\small
\begin{tabular}{|l|l|}
\hline
$~~~~{\rm parameters}~~~~$  &$~~~~~~~~~~~~~~~~~m(A_0)/g_3^2~~~~~~~~~~~~$\\
\hline
$\beta~~~\vline~~~~x~~~~\vline~~~~h~~~~~$ &$~~~{\rm measured}~$ \vline
$~~~{\rm constituent}$\\
\hline
$~9~~~~0.10~~~-0.2883$  &$~~~2.41(2)~~~~~~~~~~~~~2.74(2)~$\\
$16~~~~0.05~~~-0.2314$  &$~~~2.28(20)~~~~~~~~~~~~2.38(16)$\\
$24~~~~0.03~~~-0.1475$  &$~~~3.03(65)~~~~~~~~~~~~3.28(30)$\\
\hline
\end{tabular}
\vskip0.2truecm
Table~3: {The masses of the scalar bound state measured 
in units of $g_3^2$ compared with
the predictions of the constituent model. The measured value in the
fisrt raw was taken from Ref. \cite{philipsen1} the two other values
were taken from Ref. \cite{ours}.}
}
\end{center}
\vspace*{0.1cm}
\noindent
{\bf Acknowledgments:} This work was partly supported by the TMR 
network {\it Finite Temperature Phase Transitions in Particle Physics}, 
EU contract no. ERBFMRX-CT97-0122. The numerical work has been performed
at the HLRZ J\"ulich and the computer center of the University of Stuttgart.

\end{document}